\documentclass{article}
\usepackage{ismir,amsmath,amssymb,cite,url}
\usepackage{graphicx}
\usepackage{color}
\usepackage{subcaption}

\title{LEARNING TO TRANSCRIBE BY EAR}

\oneauthor
  {Rainer Kelz, Gerhard Widmer} {Institute of Computational Perception\\Johannes Kepler University Linz, Austria\\ {\tt rainer.kelz@jku.at}}

\newcommand\sourcecoderepo{\url{http://available.upon.publicat.ion}}
\newcommand\midifiles{\url{http://listen.ing}}

\begin{document}

\maketitle

\begin{abstract}
  Rethinking how to model polyphonic transcription formally, we frame it as a reinforcement learning task. Such a task formulation encompasses the notion of a musical agent and an environment containing an instrument as well as the sound source to be transcribed. Within this conceptual framework, the transcription process can be described as the agent interacting with the instrument in the environment, and obtaining reward by playing along with what it hears.

  Choosing from a discrete set of actions - the notes to play on its instrument - the amount of reward the agent experiences depends on which notes it plays and when. This process resembles how a human musician might approach the task of transcription, and the satisfaction she achieves by closely mimicking the sound source to transcribe on her instrument.

Following a discussion of the theoretical framework and the benefits of modelling the problem in this way, we focus our attention on several practical considerations and address the difficulties in training an agent to acceptable performance on a set of tasks with increasing difficulty. We demonstrate promising results in partially constrained environments.
\end{abstract}
\section{Introduction}\label{sec:introduction}
Polyphonic transcription is the task of extracting a set of symbolic notes from audio. Typical transcription systems, as exemplified by \cite{benetos_2013, cheng_2016, sigtia_2016, ewert_2016, hawthorne_2017} work on spectrogram frames and output a note activity vector for each frame. These note activity vectors are the input to a temporal smoothing, thresholding and decoding stage, finally leading to the construction of note objects. A note object, or symbolic note, can be described as a tuple $(t_{s}, t_{e}, p, v)$ comprised of a start point (onset) and end point (offset) in time, a symbolic pitch $p$, directly related to the fundamental frequency of the sound. Optionally, a note object might indirectly encode the volume of the note via the so called velocity $v$.

Solving a polyphonic transcription task therefore encompasses explaining a dense mixture of various sounds over time in terms of comparatively fewer and temporally much sparser distributed causes, such as the discrete decision to press a key on a musical keyboard, or pluck a string at a particular point in time, as well as choosing when to lift the finger from the key, or muting the string again.

\begin{figure}[t]
  \centering
  \includegraphics[scale=0.3]{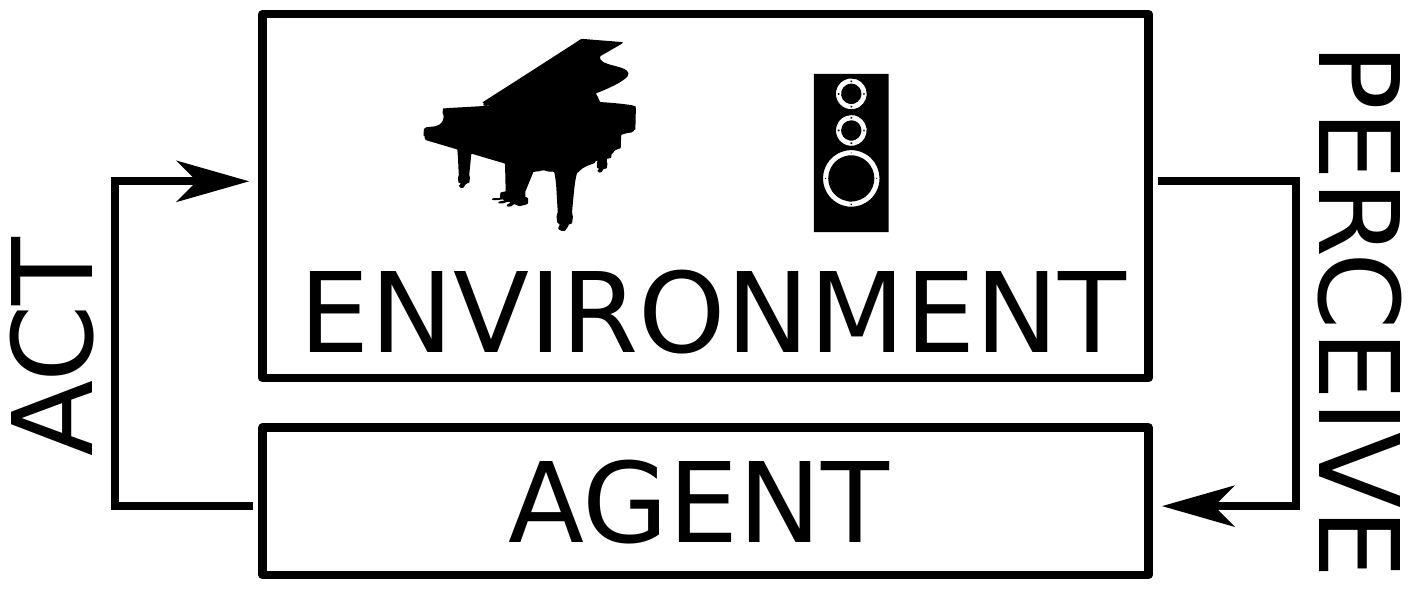}
  \caption{Sketch of the agent-environment interaction loop}
  \label{fig:sketch_agent_environment_loop}
\end{figure}

All of the above mentioned transcription systems obtain approximate solutions to the problem by solving a continuous optimization problem. The continuous optimization problem usually arises by introducing smooth relaxations. Discrete step functions are replaced by smooth nonlinearities, such as the $\tanh$ function or scaled versions thereof. Discrete performance measures, such as accuracy or precision, recall and f-measure are replaced by surrogate loss functions such as mean squared error or binary cross entropy. This necessitates having either fixed thresholds to obtain discrete decisions after training, or tuning thresholds on holdout data.

Instead of optimizing surrogate losses we opt to directly model the discrete decision making process of playing individual notes in a musical piece, together with the consequences of these individual decisions. 

We choose to model the polyphonic transcription task in a way that is reminiscent of a human listening to audio, trying to recreate what she hears by playing along on her instrument. The better the recreation produced by a musical agent, be it human or machine, the more rewarding the experience. In \cite{hainsworth_2003} an informal study on how musicians transcribe audio recordings was undertaken. The authors report that many people describe their approach roughly as \begin{quote}[...] build up a mental representation, and then play this on an instrument, sometimes along with the original audio. [...]\end{quote}

A sketch of the interaction model can be seen in \mbox{Figure \ref{fig:sketch_agent_environment_loop}}.
There are two main entities, the agent and the environment that the agent interacts with. The musical agent perceives the state of the environment, which is comprised of a few spectrogram frames of the sound source that the agent wants to reproduce, and the current state of the instrument on which to reproduce the piece. The state of the instrument is discrete and reflects which notes are sounding currently.

The reward of the agent is given by the environment and reflects the acoustic similarity between what is heard from the sound source and what is currently being played on the instrument. If what is being played is identical to what should be reproduced, the reward is maximal.

In section \ref{sec:mdps} we will introduce the necessary formalisms to model the transcription problem and solve it via reinforcement learning methods, section \ref{sec:policy_gradient_methods} will introduce actual solution methods, in section \ref{sec:environments} we will describe key aspects of the environment, and in section \ref{sec:experiments} we will discuss results from computational experiments.

\section{Markov Decision Processes}\label{sec:mdps}
We will now formally define Markov decision processes, and then cast the transcription problem as such a process. This section as a whole relies heavily on \cite{sutton_barto_2018}, chapters \mbox{1-3}. We also adhere to the notational style of \cite{sutton_barto_2018}, where uppercase letters refer to random variables and lowercase letters to their realizations.

A Markov decision process (MDP) is a collection of random variables ordered in time. Time is assumed to be discrete, and subscripts index the time dimension. An MDP is described by a tuple $(\mathcal{S}, \mathcal{A}, \mathcal{R}, \mathcal{T}, \gamma)$ consisting of a state space $\mathcal{S}$, a set of possible actions $\mathcal{A}(s)$ to take in a particular state $s$, a scalar reward function $\mathcal{R}$ that determines how much immediate reward is gained from being in a particular state $s$ at a particular point in time $t$, and a stochastic state transition function $\mathcal{T}$ in the form of a conditional probability distribution over successor states $p(s' | s, a)$, conditioned on the current state and a particular action. This distribution has the Markov property, meaning that the following equality holds:
\[
p(S_{t} | S_{t-1}, A_{t-1}) = p(S_{t} | S_{t-1}, A_{t-1}, \dots S_{0}, A_{0})
\]

The evolution of a Markov decision process over time starts with a random state $S_0$. The agent perceives the state, and decides on an action $A_0$. The environment reacts to the action $A_0$ and determines its consequences - the successor state $S_1$ and a reward $R_1$ - whereupon the cycle begins anew. This leads to trajectories through state space:

\[
S_0, A_0, R_1, S_1, A_1, R_2, S_2, A_2 \dots
\]
If these trajectories have a natural end, they are also referred to as episodes.

The behavior of the agent is governed by the policy $\pi(a | s)$. The policy is a probability distribution over actions, conditioned on the state $s$. The only goal of an agent is to behave in such a way as to maximize the expected future reward $v_{\pi}(s) = \mathbb{E}[G_t | S_t = s]$, from the current state onwards. The function $v_{\pi}(s)$ is called the state value function. It is defined in terms of the return $G_t$, which is the sum of all future reward discounted by the factor $\gamma \in [0, 1]$

\[
G_t = \sum_{k=0}^{\infty} \gamma^k R_{t+k+1}
\]

How much value a state gets assigned, depends on the behavior, the policy $\pi$. An agent has to learn a policy by interaction with the environment such that the expected future reward $v_{\pi}(\cdot)$ is maximized.
%Next to $v_{\pi}$ we can define a state action value function $q_{\pi}(s, a)$ as well, which encodes the value of taking the action $a$ in state $s$, and following the policy $\pi$ from then on.
%Learning which actions to choose and when is also called the temporal credit assignment problem.
We will now discuss modelling polyphonic transcription as an MDP in greater detail.

\subsection{Transcription as an MDP} \label{sec:transcription_as_mdp}

%% \begin{figure}[t]
%%   \centering
%%   \includegraphics[scale=0.2]{figs/agent_environment_loop.pdf}
%%   \caption{The agent-environment interaction loop}
%%   \label{fig:agent_environment_loop}
%% \end{figure}

%% \begin{figure}
%%  \centering
%%  \includegraphics[scale=0.2]{figs/state_reward_action.pdf}
%%  \caption{Illustrations of state, reward function, and actions}
%%  \label{fig:state_reward_action}
%% \end{figure}

\begin{figure}[t!]
    \centering
    \begin{subfigure}[t]{0.5\columnwidth}
        \centering
        \includegraphics[scale=0.24]{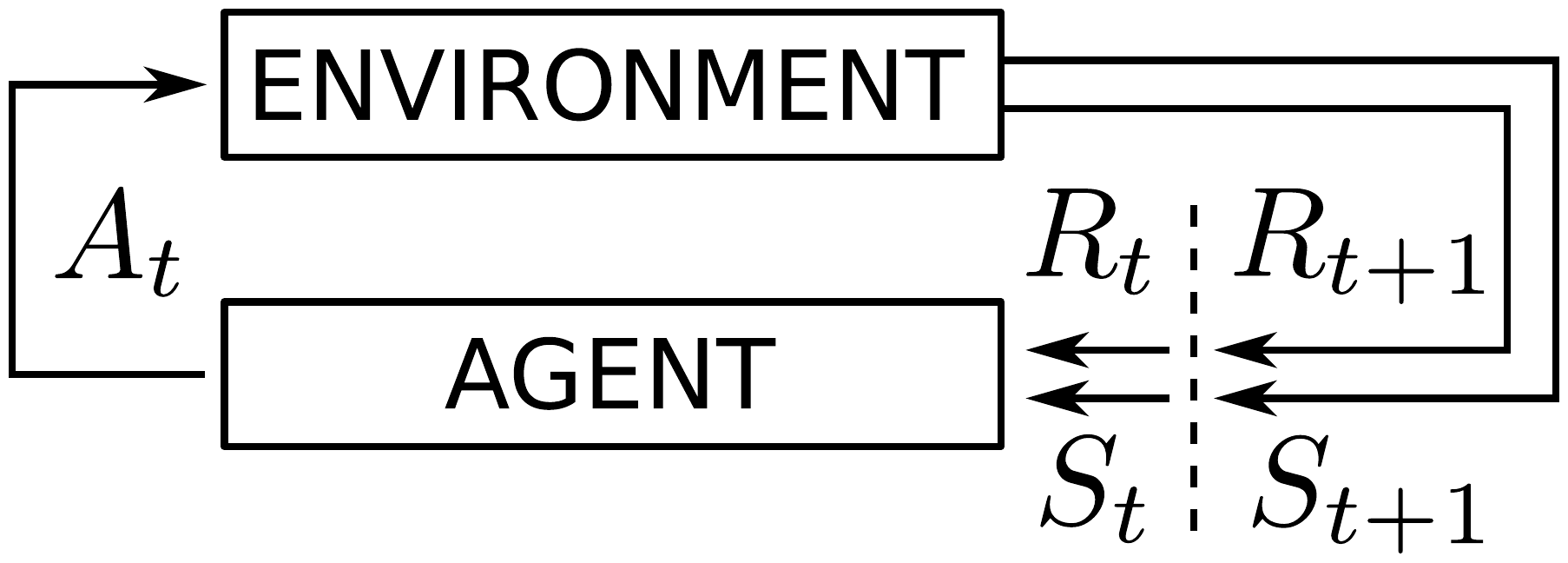}
        \caption{Interaction loop}
        \label{fig:agent_environment_loop}
    \end{subfigure}%
    ~ 
    \begin{subfigure}[t]{0.5\columnwidth}
        \centering
        \includegraphics[scale=0.24]{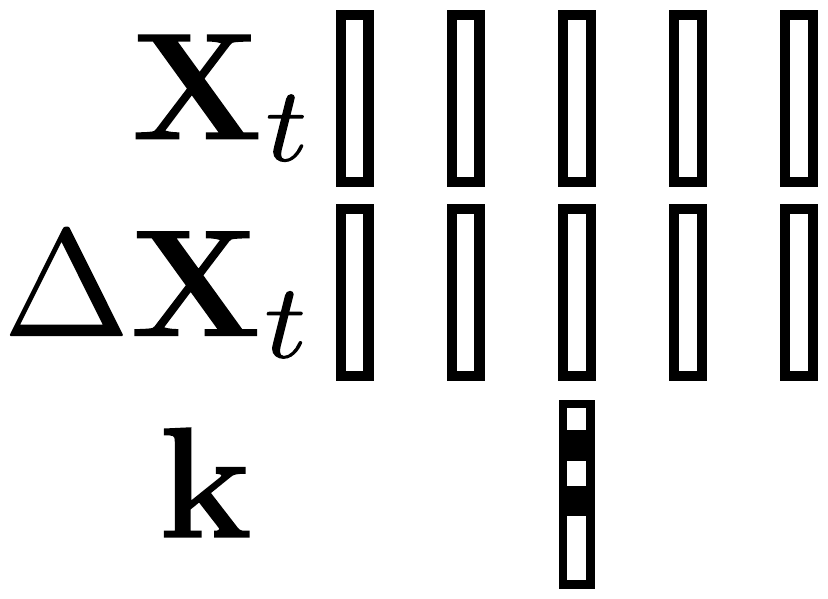}
        \caption{The definition of a state}
        \label{fig:state_only}
    \end{subfigure}
    \caption{Illustration of the agent-environment interaction loop, and how states, reward and actions look like}

\end{figure}

As mentioned in the introduction, we model transcription as the interaction of a musical agent with an environment. The agent perceives music from an ambient sound source, and plays along on its instrument. The closer the instrumental reproduction, the more reward is experienced.

 Time is assumed to be discrete, subscripts index timesteps. Interaction is strictly formalized such that the only inputs the agent has at its disposal are the current state $S_t$ and the current reward $R_t$, and the only way for the agent to interact with the environment is to decide on the current action $A_t$, as is illustrated in Figure \ref{fig:agent_environment_loop}. After time advances by one step, the agent perceives the next state $S_{t+1}$ as well as the next reward $R_{t+1}$.

 The main modelling effort lies in specifying what a state looks like, how the environment reacts to an action by transitioning to a new state and how it assigns reward. Although we investigated multiple versions of the environment with slight differences in how the state was constructed, we describe only a prototypical version here. Exact specifications of environments, agents and models to reproduce all findings are available as source code online\footnote{\sourcecoderepo}.

 We define the state as a tuple $s = (\mathbf{X}_t, \Delta \mathbf{X}_t, \mathbf{k})$. The vector $\mathbf{x}_t$ describes the current spectrogram frame for the sound source that should be transcribed. The matrix \mbox{$\mathbf{X}_t = [\mathbf{x}_{t-c}, \dots, \mathbf{x}_{t}, \dots, \mathbf{x}_{t+c}]$} consists of the last $c$ spectrogram frames, the current frame at time $t$ and $c$ future frames. The overall lag of the system with respect to the sound source to be transcribed is therefore exactly $c$ frames. Additionally, $\Delta \mathbf{X}_t$ encodes the finite difference in time direction as a low level feature sensitive to onsets. This is redundant, as this information is already contained in $\mathbf{X}_t$, but facilitates learning for the class of models under consideration in section \ref{sec:experiments}. $\mathbf{k}$ is a binary indicator vector encoding the current state of the virtual musical keyboard of the instrument. If the $k$th key is pressed, meaning that the $k$th note is currently sounding, the $k$th component of this vector is set to $1$, otherwise it is $0$. The environment knows the exact, discrete current state of the instrument because of the agent's previous actions leading up to this state, namely which keys have been pressed and released again.

The reward $R_t$ at each timestep is what implicitly defines the goal of a reinforcement learning task. It should be high if a state is desirable, and low in undesirable states. For the case of transcription, any measure appears to be reasonable that measures the similarity between the sound to be transcribed, and the sound produced by the agent's virtual instrument. We investigate multiple audio similarity measures, but defer discussion of their differences and resulting differences in agent behavior until section \ref{sec:experiments}.

The MDP formalism allows for the agent to decide on exactly one action $A_t$ in each timestep. This would exclude polyphonic instruments. To allow for polyphony, we define the actions $\mathcal{A}(s)$ available in a state as playing all possible note combinations from the set of playable notes. If the tonal range of the instrument is $K$, the number of all possible note combinations is $2^K$. This would lead to an unnecessarily large number of outputs for the model. We therefore output a binary indicator vector with $K$ components that indexes into the set of possible actions.

\subsection{Benefits}
We are convinced that this way of looking at the problem has merit. After its formal introduction we would like to emphasize some of the benefits of the MDP perspective on transcription. To summarize the previous section, the agent perceives a short spectrogram snippet (the matrix $\mathbf{X}_t$) and a low-level onset feature (the matrix $\Delta \mathbf{X}_t$) of the sound source it is supposed to transcribe, as well as the state of the musical keyboard of the virtual instrument it is playing by way of an indicator vector (the vector $\mathbf{k}$) that tells the agent which keys are currently pressed. The action an agent chooses at each time step is in the form of a binary vector as well, indicating whether to press a key or stop pressing it.

This perspective on the problem of transcription has multiple benefits. Labelled data becomes unnecessary to learn a transcription model. Supervision is replaced by interaction with a computer controllable instrument, such as a software sampler, a physical simulation of an instrument, or even an actual automatic reproducing piano, such as a Disklavier. \textit{Anything} that is controllable by MIDI messages can be used to learn to transcribe unlabelled data.

A second benefit, closely tied in with the first, is that the temporal evolution of notes is modelled implicitly. The agent can only start and stop a note, there is no leeway in the decision making process to modify the note intermediately in any way. Hence there is also no need to try and explicitly force this behavior, for example via a constraint to a continuous optimization problem, as done in \cite{ewert_2016}.

A third benefit is that in order to transcribe new, unseen material, the agent simple needs to be set loose and allowed to maximize reward. Depending on how much music it has listened to in the past, it will start from an already reasonable solution, that is then iteratively refined. As an added bonus, we obtain a better, more general recognition model during this procedure, which can then be applied to the next piece - the model improves continuously with each application.

\section{Policy Gradient Methods} \label{sec:policy_gradient_methods}
We will now discuss a particular solution method for MDPs that directly learns a policy $\pi$. This discussion relies heavily on \cite{sutton_barto_2018}, chapter 13.

Readers not familiar with (or not interested in) the details of reinforcement learning may decide to skip this section - which will describe the details of the specific learning algorithms we chose - and re-join us in section \ref{sec:environments} or \ref{sec:experiments}, where environment implementation and experimental evaluation are described.

We assume that the conditional distribution over actions $\pi(a|s)$, the policy of the musical agent, is given by a function approximator with the parameter vector $\boldsymbol{\theta}$.

The performance measure $J(\boldsymbol{\theta})$ for episodic reinforcement learning, which is to be maximized, can be defined as the value of the state the episode starts in

\[
J(\boldsymbol{\theta}) = v_{\pi_{\boldsymbol{\theta}}}(s_0)
\]

The gradient of this performance measure is

\[
\nabla J(\boldsymbol{\theta}) = \mathbb{E}_{\pi_{\boldsymbol{\theta}}} \left[ G_t \frac{\nabla_{\boldsymbol{\theta}}\pi_{\boldsymbol{\theta}}(A_t|S_t)}{\pi_{\boldsymbol{\theta}}(A_t|S_t)} \right]
\]

Which simplifies to

\[
\nabla J(\boldsymbol{\theta}) = \mathbb{E}_{\pi_{\boldsymbol{\theta}}} \left[ G_t \nabla_{\boldsymbol{\theta}}\ln\pi_{\boldsymbol{\theta}}(A_t|S_t) \right]
\]

via the identity $\frac{\nabla f}{f} = \nabla \ln f$. This expectation of the gradient can be approximated by sampling, leading to the stochastic update rule

\[
\boldsymbol{\theta} \leftarrow \boldsymbol{\theta} + \alpha G_t \nabla_{\boldsymbol{\theta}}\ln\pi_{\boldsymbol{\theta}}(A_t|S_t)
\]

also called the REINFORCE rule \cite{williams_1992}. This update rule has the simple interpretation of increasing the probability of choosing action $A_t$ in state $S_t$, proportional to the return $G_t$. In other words, as the return tells us how much cumulative reward we could achieve by being in state $S_t$ and choosing action $A_t$, and then continue to behave according to $\pi_{\boldsymbol{\theta}}$, we can use it to scale the step in parameter space that increases the probability of choosing the action $A_t$ because it helped lead to this amount of return. If we have a large return, we would like to make the action that leads to it more probable than if we only have a small return. $\alpha$ is an additional step size, or learning rate, to keep the size of the update reasonably small. For the complete derivation of this rule via the policy gradient theorem, we refer to \cite{sutton_barto_2018}, chapter 13.2. To apply the REINFORCE algorithm, the agent generates an episode using its (stochastic) policy, accumulating all gradient updates for all state-action-return triples, and then updating the parameter vector. This cycle repeats until some computational budget is used up.
%% Pseudocode for REINFORCE is written up as Algorithm \ref{alg:reinforce} for the convenience of the reader.

%% \begin{algorithm}[t]
%% \begin{algorithmic}
%% \State initialize $\boldsymbol{\theta}$ appropriately for function class of $\pi_{\boldsymbol{\theta}}$
%% \While{within computational budget}
%% \State generate $S_0, A_0, R_1, \dots, S_{T-1}, A_{T-1}, R_{T}$ by \\  \hspace{12pt} following stochastic policy $\pi_{\boldsymbol{\theta}}(\cdot | \cdot)$
%% \For{$t \,\,\mathbf{in}\,\, 0\,\,..\,\,T-1$}
%% \State $G_t \leftarrow \sum_{k=0}^{T-t}R_{t+k+1}$
%% \State $\boldsymbol{\theta} \leftarrow \boldsymbol{\theta} + \alpha G_t \nabla_{\boldsymbol{\theta}}\ln\pi_{\boldsymbol{\theta}}(A_t|S_t)$
%% \EndFor
%% \EndWhile
%% \State \Return $\boldsymbol{\theta}$
%% \end{algorithmic}
%% \caption{REINFORCE}
%% \label{alg:reinforce}
%% \end{algorithm}

\subsection{Baseline, Actor-Critic}
The REINFORCE algorithm is a Monte Carlo algorithm. The true gradient of the objective is an expectation over all possible trajectories through the state space which is approximated by sampling from all trajectories according to the current policy $\pi_{\boldsymbol{\theta}}$. This makes the gradient estimate quite noisy, as the return $G_t$ may vary wildly. To reduce the variance of this estimate, a baseline is introduced that does not change the expected value of the update, but reduces its variance. This baseline is almost always chosen to be an estimate of the true value function.
$\hat{v}_{\boldsymbol{\phi}}(\cdot) \approx v_{\pi_{\boldsymbol{\theta}}}(\cdot)$, with the estimator $\hat{v}_{\boldsymbol{\phi}}$ having a different set of parameters than the policy. Reinforcement learning schemes that use policy gradient methods in combination with value function estimation can be classified as actor-critic methods, where the policy represents the actor, and the value function estimate serves as the critic.

\subsection{A3C, A2C}
In order to increase learning stability and therefore learning speed, asynchronous actor-critic methods were investigated in \cite{mnih_2016} for deep neural network architectures. The main idea is to have multiple independent copies of agents that simultaneously explore the state space, accumulate their individual gradient estimates and asynchronously update a global set of parameters, that is distributed to the multiple independent copies periodically. This method is called asynchronous advantage actor-critic, or A3C. It was empirically determined in \cite{wu_2017} that synchronous agents and synchronous updates work just as well, and in some cases even better, which lead to the method called advantage actor-critic, abbreviated as A2C. Although mainly developed to stabilize deep neural networks, the method has an equally stabilizing and accelerating effect on shallow, almost linear models.

%% \subsection{Exploration}
%% To ensure continuous exploration, the policy $\pi_{\boldsymbol{\theta}}$ is a stochastic policy. This means that the final discrete choice for an action is made by sampling from the discrete distribution over actions, parametrized by $\boldsymbol{\theta}$.
%% To further encourage exploration of the state space, \cite{mnih_2016} report that an additional entropy term added to the policy output counteracts an early collapse to a deterministic policy. Such a collapse means that all the probability mass ends up concentrated in one discrete event. This effect is clearly visible in the reward and entropy curves discussed in section \ref{sec:experiments}, and can be detrimental to performance if it happens too early.

\section{Environments and Agents} \label{sec:environments}
%% \showthe\columnwidth
%% \uselengthunit{in}\printlength{\columnwidth}

We design OpenAI Gym \cite{brockman_2016} environments\footnote{\sourcecoderepo} to experiment with different reward formulations and determine the feasibility of learning to transcribe by ear. All the environments use a software sampler called Fluidsynth\footnote{\url{www.fluidsynth.org}} together with the soundfont ``Fluid R3 GM''\footnote{\url{http://www.musescore.org/download/fluid-soundfont.tar.gz}} for simulating the instrument that the agent plays. In addition, we also use the same soundfont to produce the input sound (the external sound source) in our preliminary experiments. Of course, assuming that the learner has access to the exact same instrument that it is trying to transcribe is a more than unrealistic assumption in practice. However, as the purpose of our initial experiments is simply to investigate if the general approach is feasible at all, we consider this a legitimate simplification. A convenient side effect is that we have access to the groundtruth, enabling us to empirically determine suitable reward functions.

\begin{figure}[t]
  \centering
  \includegraphics[width=0.8\columnwidth]{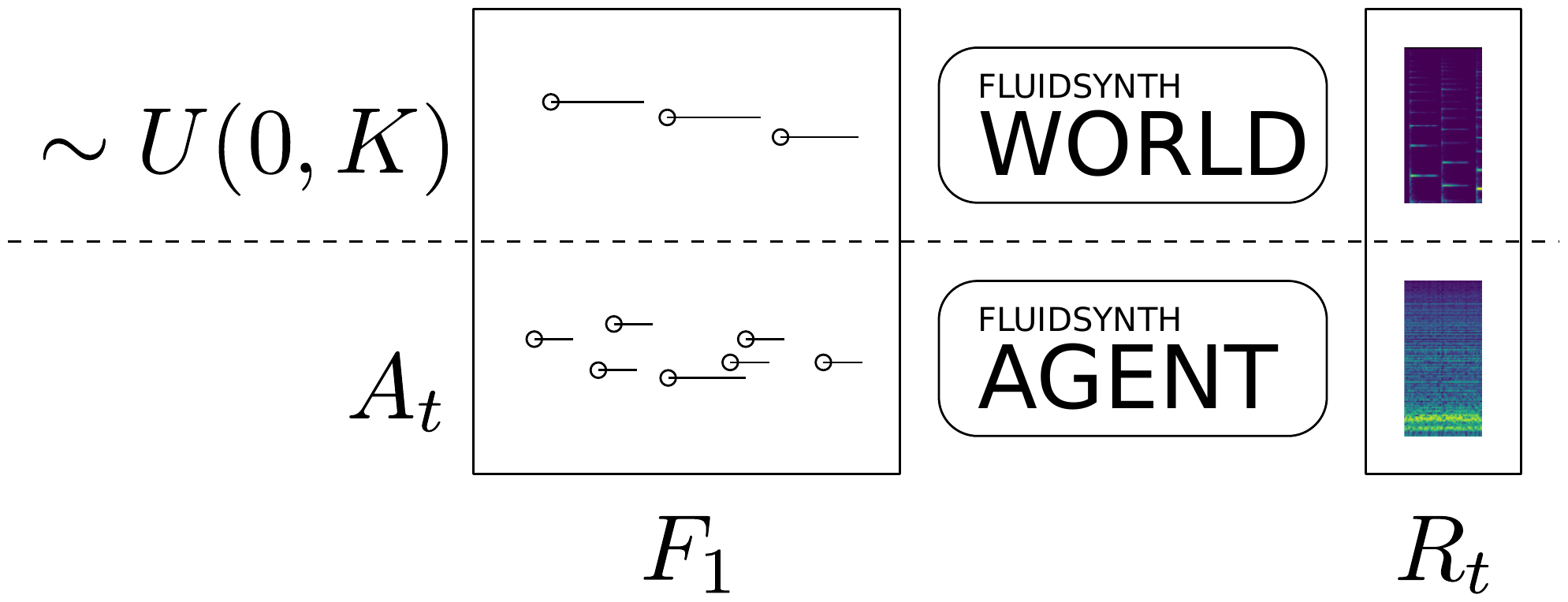}
  \caption{The inner workings of the environment. The sound source that should be transcribed by the agent plays random notes, the agent plays notes $A_t$ according to its current policy, and both MIDI streams are rendered to audio separately. An audio similarity measure is used on the two audio streams to define the reward $R_t$. In the feasibility study, the $F_1$ score for the two MIDI streams is used to determine a suitable definition for $R_t$.}
  \label{fig:environment}
\end{figure}

The prototypical environment we use for our experiments is split into two parts. The ``world'' part, and the ``agent'' part, as illustrated in Figure \ref{fig:environment}. The tonal range of both sound sources is constrained to one octave \mbox{(C4 - B4)} and one instrument to learn from and play with (a piano). The musical content of the ``world'' sound source is chosen uniformly at random with varying degrees of polyphony.

This rather simple environment already provides us with a lot of insight into the nature of the problem, what kind of policies are learned and how the solution methods behave during training.

\begin{figure}[t]
  \centering
  \includegraphics[width=0.8\columnwidth]{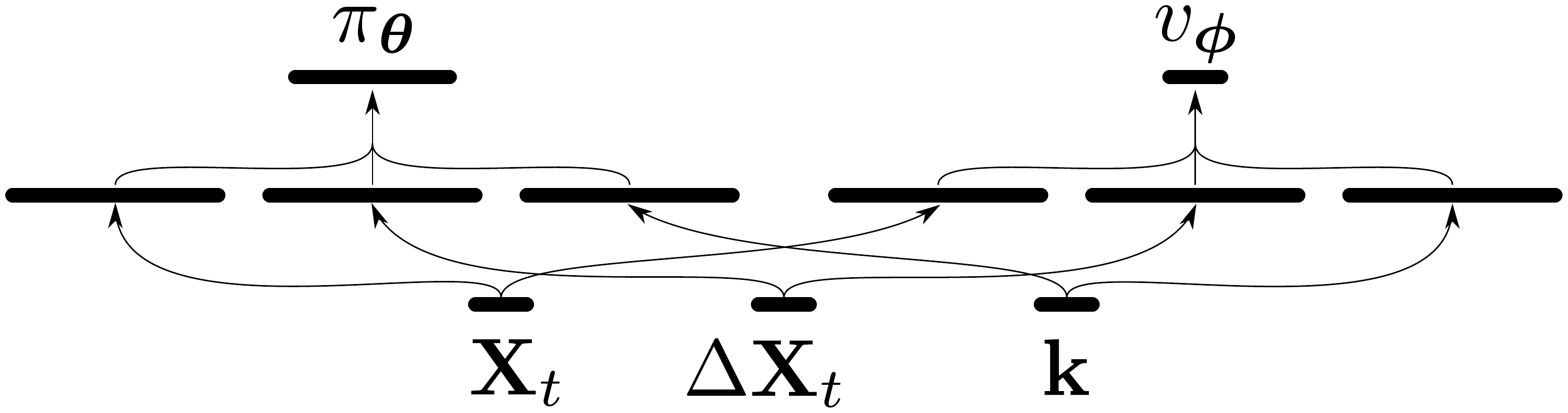}
  \caption{The structure of the model used for learning the parametrized policy $\pi_{\boldsymbol{\theta}}$ and the parametrized value function $v_{\boldsymbol{\phi}}$}
  \label{fig:model}
\end{figure}

For the sake of simplicity, we choose all policies $\pi_{\boldsymbol{\theta}}(a|s)$ to be parametrized by shallow, linear neural networks, as illustrated in Figure \ref{fig:model} with each arrow representing the action of a linear layer. Horizontal black bars are the activations after a layer. Arrows converging indicate that the intermediate results are concatenated into a larger vector. There are two versions of the model, a monophonic and a polyphonic version. The monophonic version applies a softmax function over a vector that has as many components as the tonal range of the instrument it is playing, and an additional component to encode that it does nothing. The polyphonic version with the factorized policy outputs an indicator vector with as many components as the tonal range of the instrument it is playing.

\section{Experiments} \label{sec:experiments}
For training an agent in the reinforcement learning scenario as described, we do not need any label information. Having a virtual instrument that sounds moderately similar to what should be transcribed is sufficient. For experimental validation purposes we would still like to use the performance measures we are accustomed to, such as note-overlap, or framewise $F_1$ score. It is not a priori clear that the reward function we choose correlates with note-overlap or $F_1$ score, so we try and empirically determine a suitable reward function first.

We optimize our models not directly with the \mbox{REINFORCE} update rule, instead opting for the Adam \cite{kingma_2014} update rule which is much more robust against noise and unfortunate hyperparameter choices. We consider all the following as hyperparameters: discount factor $\gamma$, learning rate $\alpha$, mean momentum term $\beta_1$, policy entropy term weight $\eta$, and number of hidden units. We would like to emphasize again that the exact configurations are available as source code\footnote{\sourcecoderepo}.

\subsection{Monophonic World - Monophonic Agent}
We start by evaluating multiple audio similarity measures qualitatively on a simplified transcription task by observing the correlation between reward and $F_1$ score. For this simplified task, the environment renders random monophonic melodies and a monophonic agent needs to learn to transcribe them.

The parameterization of the agent is the same across all runs, and the same hyperparameters are tested for all different reward formulations. The agent can only maximize reward, and selection of the best among multiple agents is based on the reward as well. If the reward function does not correlate with the $F_1$ score, it is ineffective. We always show a run that achieved high average reward, together with a low performing run, where low is always relative to all other runs.

\begin{figure}[t]
  \centering
  \includegraphics[scale=1.0]{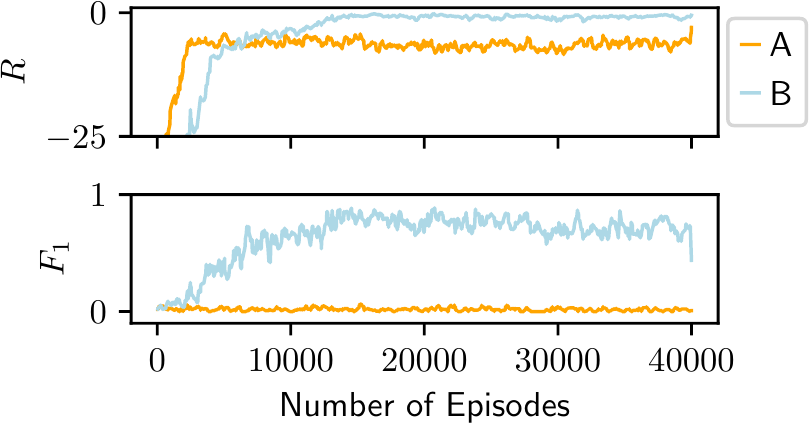}
  \caption{Qualitative comparison of the reward definition in Eq. \ref{eq:minkowski} with the inaccessible $F_1$ score for two runs with different hyperparameter configuration.}
  \label{fig:minkowski_correlation}
\end{figure}

All reward definitions are a similarity measure between the current spectrogram frame of the world audio ($\mathbf{x}_t$) and the current spectrogram frame of the agent audio ($\mathbf{y}_t$) in the current timestep. We start with an obvious choice for a similarity measure, the negative $L_2$ norm:

\begin{equation}
  R_t = -\|\mathbf{x}_t - \mathbf{y}_t\|
  \label{eq:minkowski}
\end{equation}

A qualitative comparison between two runs, both using this reward definition but having otherwise different hyperparameters is depicted in Figure \ref{fig:minkowski_correlation}. We observe that for configuration ``A'' the reward is uncorrelated with $F_1$ for a short phase at the beginning of the optimization process. For configuration ``B'', the degree of correlation with $F_1$ score changes while learning progresses. Although this is only one example, it is representative of how the bigger picture looks like. We conclude that this reward formulation might be useful eventually, if there is a way to calibrate it so that the mean over runs with configuration ``A'' becomes a lower bound on the reward. The runs that accumulated the most reward produced a transcription that was qualitatively acceptable, meaning that all note onsets were transcribed nearly perfectly, however the sustain and release phases of most notes were mostly neglected. This was expected, as both the $L_2$ and the $L_1$ norm are very sensitive to the actual signal power, which concentrates at the onsets for pitched percussive instruments such as the piano. The use of this reward definition leads to high precision, but low recall in general. We note that the negative $L_1$ norm leads to almost identical solutions.

We ran the exact same hyperparameter configurations with the cosine similarity as the reward signal next:

\begin{equation}
    R_t = \frac{\mathbf{x}_t \cdot \mathbf{y}_t}{\|\mathbf{x}_t\| \|\mathbf{y}_t\|}
    \label{eq:cos}
\end{equation}

\begin{figure}[t]
  \centering
  \includegraphics[scale=1.0]{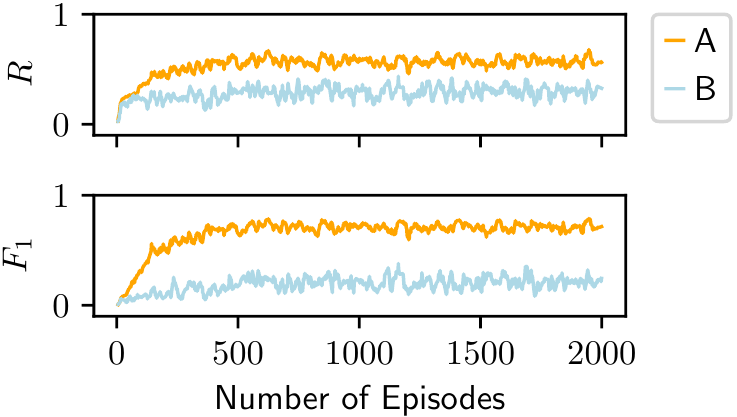}
  \caption{Qualitative comparison of the reward definition in Eq. \ref{eq:cos} with the inaccessible $F_1$ score for two runs with different hyperparameter configuration.}
  \label{fig:cos_correlation}
\end{figure}

All reward curves correlated nicely with the framewise $F_1$ score and training was less sensitive to hyperparameters overall. The transcriptions obtained were qualitatively worse than the ones obtained with $L_2$ or $L_1$ as the reward. This is because the cosine similarity is close to $0$ when either one of $\mathbf{x}_t$ or $\mathbf{y}_t$ are close to zero, but the respective other is not. The use of this reward definition leads to high recall, but low precision in general, because playing random notes while no external sound source is heard is not discouraged.

In an attempt to combine these two similarity measures, we choose the Hellinger distance (Eq. \ref{eq:hellinger}) measure as a replacement for the euclidean distance, with $\boldsymbol{1}$ denoting the vector of all ones. It is defined on discrete probability distributions and its range lies within the interval $[0, 1]$, which makes it ideal for combination with the cosine similarity. The two similarities are supposed to compensate their respective shortcomings.

\begin{align}
  H(\mathbf{u}, \mathbf{v}) & = \frac{1}{\sqrt{2}} \left\|\sqrt{\frac{\mathbf{u}}{\boldsymbol{1} \cdot \mathbf{u}}} - \sqrt{\frac{\mathbf{v}}{\boldsymbol{1} \cdot \mathbf{v}}} \right\| \label{eq:hellinger} \\
  C(\mathbf{u}, \mathbf{v}) & = \frac{\mathbf{u} \cdot \mathbf{v}}{\|\mathbf{u}\| \|\mathbf{v}\|} \\
    R_t & = \max \left[C(\mathbf{x}_t, \mathbf{y}_t), 1 - H(\mathbf{x}_t, \mathbf{y}_t)\right] \label{eq:cnr_hell}
\end{align}

\begin{figure}[t]
  \centering
  \includegraphics[scale=1.0]{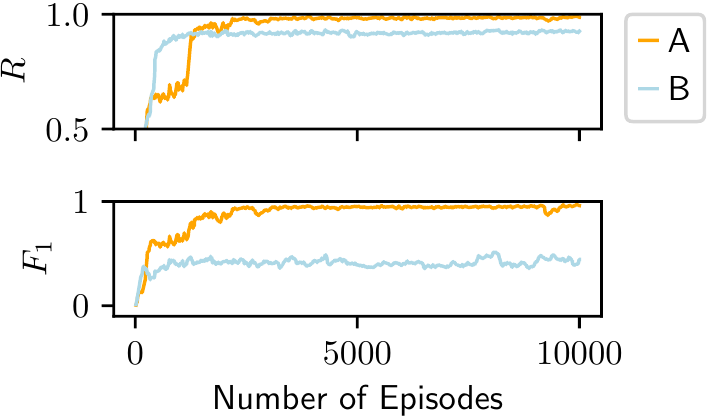}
  \caption{Qualitative comparison of the reward definition in Eq. \ref{eq:cnr_hell} with the inaccessible $F_1$ score for two runs with different hyperparameter configuration.}
  \label{fig:cnr_hell_correlation}
\end{figure}

As we can observe in Figure \ref{fig:cnr_hell_correlation}, the combined reward definition has a similar problem as the negative $L_2$ norm, with bad correlation during the initial phase of optimization, and a seemingly similar need for calibration. However, as we can already see from the $F_1$ plot, the transcription result is nearly perfect, with an average framewise $F_1$ score close to $0.99$ and good correlation between reward and $F_1$ score, once the initial optimization phase is over.

\subsection{Unheard Melody - Unheard Instrument}
\begin{figure}[t]
  \centering
  \includegraphics[scale=1.0]{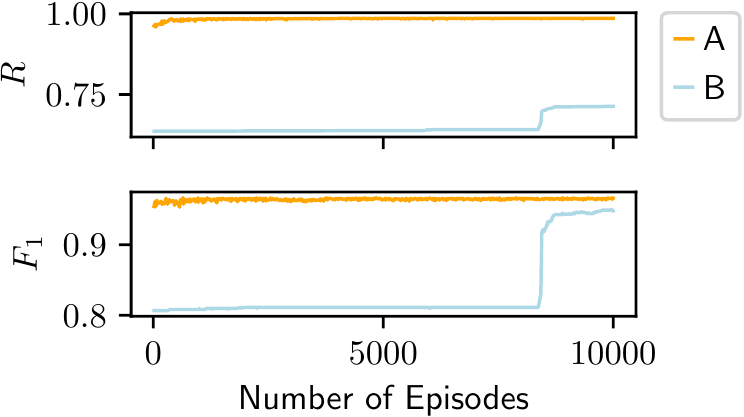}
  \caption{Adapting to an unheard instrument, playing an unheard melody. ``A'' is the performance for instruments the agent was initially trained on, ``B'' is the performance while it adapts to new, unheard instruments.}
  \label{fig:twinkle_twinkle_little_guitar}
\end{figure}

Before we move on to discuss polyphony, we would like to evaluate the transcription agent in a small informal test. To that end, we restrict the environment to only play ``Twinkle Twinkle Little Star'' with a different instrument that the agent has never heard before - a guitar. The agent under test was obtained by transcribing random notes played on one instrument for a few thousand episodes, and with the reward definition in Eq. \ref{eq:cnr_hell}. We simultaneously let the agent transcribe the melody and adapt its recognition model to try and become a better transcription agent. The instrument the agent has at its disposal to play along is still the piano it originally learned to transcribe with. As we can see in \mbox{Figure \ref{fig:twinkle_twinkle_little_guitar}}, both reward  and $F_1$ score are much lower for unknown acoustic conditions ``B'' in the beginning, than for known instruments, as depicted by ``A''. We can detect only a slight upwards trend in the initial phase for both reward and $F_1$ score - the agent improves its performance completely unsupervised, and successfully adapts to new environmental conditions. More importantly, we can observe that the agent's performance does not deteriorate, even though the instrument with which it plays along is dissimilar to what it is supposed to transcribe. After a while the agent's performance make a large jump, coming close to the transcription performance for known acoustics. Transcriptions in form of MIDI files can be listened to online\footnote{\midifiles}.

\subsection{Monophonic World - Polyphonic Agent}
\begin{figure}[t]
  \centering
  \includegraphics[scale=0.7]{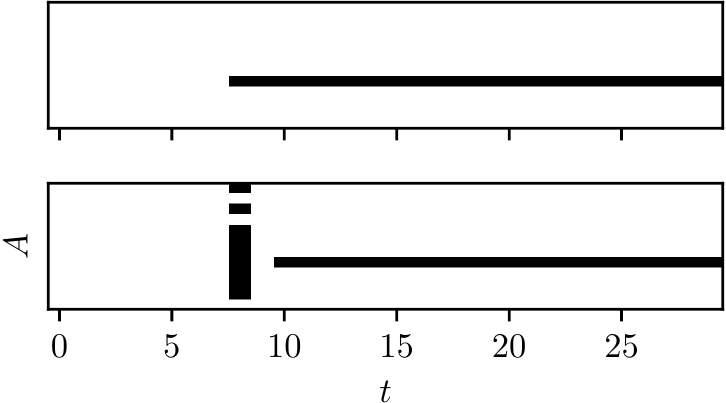}
  \caption{A representative example for peculiarities in the learned policy - clumped up key presses around an onset and subsequent recovery.}
  \label{fig:mopo_errors}
\end{figure}

In the monophonic scenario, the agent had to select among $|\mathcal{A}(s)| = 12 + 1$ actions ($12$ notes, and one do-nothing action). In this scenario the agent can select among $2^{12} = 4096$ actions at each timestep. This is a problem that is considerably more difficult. In Figure \ref{fig:mopo_errors} we can see some prototypical errors that occur during and after learning over a wide range of hyperparameters. Multiple keys ``clump'' together around onsets, due to an originally stochastic policy that has collapsed too early to a deterministic one, and hence cannot recover through exploring more rewarding states and actions.

\subsection{Polyphonic World - Polyphonic Agent}
In this more realistic scenario, where both sound sources are allowed to be polyphonic, all results indicate that the difficulties discussed for the monophonic - polyphonic scenario are also present, and much worse. An agent starting from scratch in this scenario will very likely take a very long time to learn a useable policy. To shorten this time, we already experimented with agents pretrained in a supervised fashion. These are surprisingly hard to get right, as the policies obtained by pretraining are almost deterministic and hence do not explore and adapt to new scenarios easily. There is great potential for methods that can learn from imperfect demonstrations, such as \cite{haarnoja_2017}.

\section{Conclusion}
We have argued in favor of viewing the transcription problem as a reinforcement learning problem, due to various benefits: labelled data is unnecessary, a computer controllable instrument suffices. Decisions made are discrete and symbolic note objects can be directly recovered without imposing any additional constraints or choosing thresholds. An agent can further improve its recognition model while it is transcribing unseen pieces.

We have empirically and qualitatively determined a suitable similarity measure to use as a reward signal and tested the limits of the default approach to policy learning in a large action space.

We believe that there is great potential in this modelling approach, and several salient problems to be solved that are also of relevance for the wider field of reinforcement learning, such as being able to cope with large action spaces.

As an alternative, which we left for immediate future work, one can take a step back and find an alternative formulation of the MDP. Instead of asking the agent at each timestep to decide on exactly one combination of notes (simultaenous keypresses) out of $2^K$ possible combinations, it might be an easier decision, if it were spread out over multiple timesteps (sequential keypresses). The cardinality of the action space would reduce to a much more manageable $K+1$ actions ($K$ notes and an additional action \textit{advance-real-time}). The agent would then need to decide when to press keys, and when to ``advance in real time''. If we define the hopsize of the STFT as ``one unit of real time'', this would subdivide each real time step into a variable number of ``virtual time steps'', during which the agent sequentially decides on which key to press, one at a time.

%% \begin{figure}[p]
%%  \centering
%%  \includegraphics[width=\columnwidth]{figs/Twinkle_Twinkle_Little_Star.png}
%%  \caption{Score for ``Twinkle Twinkle Little Star'' \cite{twinkle}}
%%  \label{fig:twinkle}
%% \end{figure}
%% \fi
%%%%%%%%%%%%%%%%%%%%%%%%%%%%%%%%%%%%%%%%%%%%%%%%%%%%%%%%%%%%%%%%%%%%%%%%%%%%

\bibliography{master}

\begin{thebibliography}{10}

\bibitem{benetos_2013}
Emmanouil Benetos and Simon Dixon.
\newblock {Multiple-Instrument Polyphonic Music Transcription using a
  Temporally Constrained Shift-Invariant Model}.
\newblock {\em The Journal of the Acoustical Society of America},
  133(3):1727--1741, 2013.

\bibitem{brockman_2016}
Greg Brockman, Vicki Cheung, Ludwig Pettersson, Jonas Schneider, John Schulman,
  Jie Tang, and Wojciech Zaremba.
\newblock {OpenAI Gym}.
\newblock {\em CoRR}, abs/1606.01540, 2016.

\bibitem{cheng_2016}
Tian Cheng, Matthias Mauch, Emmanouil Benetos, and Simon Dixon.
\newblock {An Attack/Decay Model for Piano Transcription}.
\newblock In {\em Proceedings of the 17th International Society for Music
  Information Retrieval Conference, {ISMIR} 2016, New York City, United States,
  August 7-11, 2016}, pages 584--590, 2016.

\bibitem{ewert_2016}
Sebastian Ewert and Mark~B. Sandler.
\newblock {Piano Transcription in the Studio Using an Extensible Alternating
  Directions Framework}.
\newblock {\em {IEEE/ACM} Trans. Audio, Speech {\&} Language Processing},
  24(11):1983--1997, 2016.

\bibitem{haarnoja_2017}
Tuomas Haarnoja, Haoran Tang, Pieter Abbeel, and Sergey Levine.
\newblock Reinforcement learning with deep energy-based policies.
\newblock In {\em Proceedings of the 34th International Conference on Machine
  Learning, {ICML} 2017, Sydney, NSW, Australia, 6-11 August 2017}, pages
  1352--1361, 2017.

\bibitem{hainsworth_2003}
Stephen~W Hainsworth and Malcolm~D Macleod.
\newblock {The Automated Music Transcription Problem}.
\newblock {\em Technical report}, pages 1--23, 2003.

\bibitem{hawthorne_2017}
Curtis Hawthorne, Erich Elsen, Jialin Song, Adam Roberts, Ian Simon, Colin
  Raffel, Jesse Engel, Sageev Oore, and Douglas Eck.
\newblock {Onsets and Frames: Dual-Objective Piano Transcription}.
\newblock {\em CoRR}, abs/1710.11153, 2017.

\bibitem{kingma_2014}
Diederik~P. Kingma and Jimmy Ba.
\newblock Adam: {A} method for stochastic optimization.
\newblock {\em CoRR}, abs/1412.6980, 2014.

\bibitem{mnih_2016}
Volodymyr Mnih, Adri{\`{a}}~Puigdom{\`{e}}nech Badia, Mehdi Mirza, Alex Graves,
  Timothy~P. Lillicrap, Tim Harley, David Silver, and Koray Kavukcuoglu.
\newblock Asynchronous methods for deep reinforcement learning.
\newblock In {\em Proceedings of the 33nd International Conference on Machine
  Learning, {ICML} 2016, New York City, NY, USA, June 19-24, 2016}, pages
  1928--1937, 2016.

\bibitem{sigtia_2016}
Siddharth Sigtia, Emmanouil Benetos, and Simon Dixon.
\newblock {An End-to-End Neural Network for Polyphonic Piano Music
  Transcription}.
\newblock {\em {IEEE/ACM} Trans. Audio, Speech {\&} Language Processing},
  24(5):927--939, 2016.

\bibitem{sutton_barto_2018}
Richard~S. Sutton and Andrew~G. Barto.
\newblock {\em {Reinforcement Learning - An Introduction}}.
\newblock Adaptive computation and machine learning. {MIT} Press, 2017.

\bibitem{williams_1992}
Ronald~J. Williams.
\newblock {Simple Statistical Gradient-Following Algorithms for Connectionist
  Reinforcement Learning}.
\newblock {\em Machine Learning}, 8:229--256, 1992.

\bibitem{wu_2017}
Yuhuai Wu, Elman Mansimov, Shun Liao, Roger~B. Grosse, and Jimmy Ba.
\newblock Scalable trust-region method for deep reinforcement learning using
  kronecker-factored approximation.
\newblock {\em CoRR}, abs/1708.05144, 2017.

\end{thebibliography}

\end{document}